\documentstyle[sprocl]{article}

\begin{document}
 
\title 
{{}FROM THE DE DONDER-WEYL HAMILTONIAN FORMALISM TO  QUANTIZATION 
OF GRAVITY\footnote{ 
submitted June 29, 1998. 
To appear in {\em Proc. Int. Sem. on Math. Cosmology, } 
Potsdam 
1998,   
 M. Rainer and H.-J. Schmidt eds. (World Sci., Singapore,1998)  } 
} 

\author{I. V. KANATCHIKOV  } 

\address{Laboratory of Analytical Mechanics and Field Theory \\
Institute of Fundamental Technological Research \\
Polish Academy of Sciences \\
\'Swi\c etokrzyska 21, Warsaw PL-00-049, Poland\\
E-mail: ikanat@ippt.gov.pl } 
\vspace*{-18mm} 
\maketitle 
\abstracts{ An approach to quantization of fields and gravity 
based on the De Donder-Weyl covariant Hamiltonian formalism 
is outlined. It leads to a hypercomplex extension of quantum 
mechanics in which the algebra of complex numbers is replaced by 
the space-time Clifford algebra 
and all space-time variables enter on equal footing. 
A covariant 
hypercomplex 
analogue of the Schr\"odinger 
equation is formulated. 
Elements of  quantization of General Relativity 
within the present framework are sketched.  
  }


\newcommand{\beq}{\begin{equation}}
\newcommand{\eeq}{\end{equation}}
\newcommand{\beqa}{\begin{eqnarray}}
\newcommand{\eeqa}{\end{eqnarray}}
\newcommand{\nn}{\nonumber}

\newcommand{\half}{\frac{1}{2}}

\newcommand{\xt}{\tilde{X}}

\newcommand{\uind}[2]{^{#1_1 \, ... \, #1_{#2}} }
\newcommand{\lind}[2]{_{#1_1 \, ... \, #1_{#2}} }
\newcommand{\com}[2]{[#1,#2]_{-}} 
\newcommand{\acom}[2]{[#1,#2]_{+}} 
\newcommand{\compm}[2]{[#1,#2]_{\pm}}

\newcommand{\lie}[1]{\pounds_{#1}}
\newcommand{\co}{\circ}
\newcommand{\sgn}[1]{(-1)^{#1}}
\newcommand{\lbr}[2]{ [ \hspace*{-1.5pt} [ #1 , #2 ] \hspace*{-1.5pt} ] }
\newcommand{\lbrpm}[2]{ [ \hspace*{-1.5pt} [ #1 , #2 ] \hspace*{-1.5pt}
 ]_{\pm} }
\newcommand{\lbrp}[2]{ [ \hspace*{-1.5pt} [ #1 , #2 ] \hspace*{-1.5pt} ]_+ }
\newcommand{\lbrm}[2]{ [ \hspace*{-1.5pt} [ #1 , #2 ] \hspace*{-1.5pt} ]_- }
\newcommand{\pbr}[2]{ \{ \hspace*{-2.2pt} [ #1 , #2 ] \hspace*{-2.55pt} \} }
\newcommand{\we}{\wedge}
\newcommand{\dv}{d^V}
\newcommand{\nbrpq}[2]{\nbr{\xxi{#1}{1}}{\xxi{#2}{2}}}
\newcommand{\lieni}[2]{$\pounds$${}_{\stackrel{#1}{X}_{#2}}$  }

\newcommand{\rbox}[2]{\raisebox{#1}{#2}}
\newcommand{\xx}[1]{\raisebox{1pt}{$\stackrel{#1}{X}$}}
\newcommand{\xxi}[2]{\raisebox{1pt}{$\stackrel{#1}{X}$$_{#2}$}}
\newcommand{\ff}[1]{\raisebox{1pt}{$\stackrel{#1}{F}$}}
\newcommand{\dd}[1]{\raisebox{1pt}{$\stackrel{#1}{D}$}}
\newcommand{\nbr}[2]{{\bf[}#1 , #2{\bf ]}}
\newcommand{\der}{\partial}
\newcommand{\oo}{$\Omega$}
\newcommand{\Om}{\Omega}
\newcommand{\om}{\omega}
\newcommand{\eps}{\epsilon}
\newcommand{\si}{\sigma}
\newcommand{\Lm}{\bigwedge^*}

\newcommand{\inn}{\hspace*{2pt}\raisebox{-1pt}{\rule{6pt}{.3pt}\hspace*
{0pt}\rule{.3pt}{8pt}\hspace*{3pt}}}
\newcommand{\sro}{Schr\"{o}dinger\ }
\newcommand{\bm}{\boldmath}
\newcommand{\vol}{\omega}
\newcommand{\dvol}[1]{\der_{#1}\inn \vol}

\newcommand{\bd}{\mbox{\bf d}}
\newcommand{\bder}{\mbox{\bm $\der$}}
\newcommand{\bI}{\mbox{\bm $I$}}

\newcommand{\be}{\beta} 
\newcommand{\ga}{\gamma} 
\newcommand{\de}{\delta} 
\newcommand{\Ga}{\Gamma} 
\newcommand{\gmu}{\gamma^\mu}
\newcommand{\gnu}{\gamma^\nu}
\newcommand{\ka}{\kappa}
\newcommand{\hka}{\hbar \kappa}
\newcommand{\al}{\alpha}
\newcommand{\lapl}{\bigtriangleup}
\newcommand{\psib}{\overline{\psi}}
\newcommand{\Psib}{\overline{\Psi}}
\newcommand{\derts}{\stackrel{\leftrightarrow}{\der}}
\newcommand{\what}[1]{\widehat{#1}}

\newcommand{\bx}{{\bf x}}
\newcommand{\bk}{{\bf k}}
\newcommand{\bq}{{\bf q}}

\newcommand{\omk}{\omega_{\bf k}}


\vspace*{-98mm}
\hbox to 4.75truein{
\footnotesize\it 
\hfil \hbox to 0 truecm{\hss 
\normalsize\rm gr-qc/9810076} 
                                   }

\vspace*{90mm}

\section{Introduction}

A synthesis of  Quantum Theory and  General Relativity remains 
one of the  fundamental problems in theoretical physics. 
Various  approaches to the problem have been proposed 
which vary 
{}from  ideas of the  revision of  our basic concepts 
about  space-time 
or  quantum theory  
to the latest  developments in string (or M-) theory 
and 
quantum gravity. 
The problem of  {\em quantization} of gravity 
represents only  one of 
the attempts to reach the above mentioned  
synthesis\cite{isham,ashtekar}.  
Again, there exist several approaches 
to quantization of gravity,   
of which the canonical quantization 
can be considered 
as  the most straightforward one.  
Usually the canonical quantization is  preceded by 
the Hamiltonian formulation. The Poisson(-Dirac) brackets found 
within the  
latter give rise to the commutation relations 
of quantum operators. The scheme can be applied, at least in 
principle, 
to General Relativity 
constituting the basis of the contemporary 
approaches to quantum gravity, 
such as the Wheeler-De Witt superspace formulation  
or the Ashtekar non-pertubative quantization program. 
However, the feature of the Hamiltonian 
formalism that a time dimension (or a foliation of space-like 
hypersurfaces) 
has to be specified prior to construction of  the  Hamiltonian formalism   
seems to be  a rather unnatural restriction in the context of 
 quantum theory of gravity.  In fact, 
the latter is  expected to imply intricate Planck-scale 
fluctuations of  topology and even of a signature of the space-time 
which are 
hardly compatible 
with the possibility of 
specifying 
a natural time parameter. 
The widely discussed ``Issue of Time'' in quantum gravity 
and cosmology\cite{isham} 
may  be viewed as another manifestation 
of conceptual problems 
which we face 
while  applying 
traditional methods 
to the realm of quantum gravity. 
In view of the above,  applicability of 
the standard Hamiltonian formalism    
to the problem of  quantization of gravity can be questioned. 

As an attempt to circumvent,  
or at least to shed new light on 
the above mentioned difficulties, 
we will discuss in what follows 
an approach to the 
quantization of fields and gravity which is based on 
a 
different 
generalization of the Hamiltonian formalism to field theory.   
This generalization,  
which is manifestly covariant and does not prerequire 
 singling out of a time dimension, 
has been  known in the 
calculus of variations 
as the De Donder-Weyl (DW) theory since the thirties,  
although its applications 
in physics have been rather rare\cite{gimm,rund,ikanat0}.   

The idea 
of the DW Hamiltonian formulation 
is that for a Lagrangian 
density  
$L=L(y^a, \der_\mu y^a, x^\nu)$,  
where $x^\nu$ are space-time coordinates $(\mu,\nu=1,...,n)$,  
$y^a $ are field variables  
$(a,b=1,...,m)$,  
and $ \der_\mu y^a $ 
are   their space-time derivatives,  
we can define Hamiltonian-like variables 
$p_a^\mu:=\der L/ \der (\der_\mu y^a)$ (polymomenta), 
and $H:= \der_\mu y^a p_a^\mu - L $  (DW Hamiltonian),  
such that the Euler-Lagrange field 
equations take the form 
of  DW Hamiltonian equations 
\beq
\der_\mu y^a = \der H / \der p^\mu_a, 
\quad \der_\mu p^\mu_a = - \der H/ \der y^a .  
\eeq 
These equations  can be viewed as a field theoretic 
(multi-parameter, or ``multi-time'')  
generalization of the Hamilton canonical equations. 
If so, it is quite natural to try to construct a quantization 
scheme in field theory which 
is related to this formulation,  
much like  
the standard canonical  quantization is 
related to the Hamiltonian formalism in mechanics.

We proceed as follows. In Sec. 2 we 
outline elements of  the quantization 
based on the Poisson brackets on differential forms  
which 
appear 
within the DW formulation\cite{ikanat0,bial96}.   
In particular, we put forward a covariant generalization of 
 the Schr\"odinger equation to the present context. 
In Sec. 3 we extend this scheme to 
 curved space-time. 
In Sec. 4 an application 
to the problem of quantization of General Relativity 
is sketched 
and 
a corresponding  generalization of the 
Schr\"odinger equation is presented.  
Discussion is found in Sec. 5.

\section{Elements of the quantization based on the DW theory} 

We have shown earlier 
that the DW Hamiltonian formulation has 
an analogue of the Poisson bracket operation 
defined on (horizontal) 
differential forms of various degrees $p$, $\,0\leq p \leq n\,$    
($n$ is  the dimension of  the underlying spacetime manifold),    
 $F=\frac{1}{p!}F\lind{\mu}{p}(y^a,p_a^\mu,x^\nu)
dx^{\mu_1}\we ... \we dx^{\mu_p}$,  
which play the role of dynamical variables\cite{ikanat0,bial96}. 
This bracket possesses (graded) analogues of the algebraic 
properties of the usual Poisson bracket, the existence  of 
which is crucial for quantization. 
More specifically, the bracket on differential forms in 
DW theory leads to generalizations of 
a so-called Gerstenhaber algebra.  
This bracket  
also enables us to identify the canonically conjugate 
variables and to write the DW Hamiltonian field equations (1) 
in Poisson bracket formulation. The latter demonstrates that the bracket 
of a form with the DW Hamiltonian is related to the (total) exterior 
differentiation of the form. This observation will be 
important below 
for our conjecture about 
an analogue of the 
Schr\"odinger equation.

For our present purposes it is sufficient to use 
canonical brackets in the subalgebra of forms of degree 
$0$ and $(n-1)$. 
Using the notation 
$\om_\mu:= (-1)^{(\mu-1)} 
dx^1\we ... \we \what{dx^\mu}\we ... \we dx^n$ 
those are given by\cite{ikanat0}   
 $$\pbr{p_a^\mu\omega_\mu}{y^b}
= 
\delta^b_a , \quad 
\pbr{p_a^\mu\omega_\mu}{y^b\omega_\nu}
=
\delta^b_a\omega_\nu, \quad 
\pbr{p_a^\mu}{y^b\omega_\nu}
= 
\delta^b_a\delta^\mu_\nu , 
\refstepcounter{equation} 
\eqno {  (\theequation a,b,c) } 
$$
with  other  brackets vanishing.   
By quantizing according to the Dirac correspondence rule, 
we conclude {}from the commutator 
corresponding to  (2a) 
that  
$\what{p_a^\mu\om_\mu} = i \der_a$  is   
the operator of partial 
differentiation with respect to the field variables. 
We argued\cite{qs96,bial97} that  operators 
$\what{p}{}_a^\mu$  and 
$\what{\om}_\mu$ can be represented by means of  
Clifford imaginary units (Dirac matrices)
$\gamma_\mu \gamma_\nu + \gamma_\nu \gamma_\mu = 2\eta_{\mu\nu},
$\footnote{A slightly 
different representation\cite{qs96}: $p_a^\mu=i\hbar\kappa\gamma^\mu\gamma\der_a$ 
with $\gamma \sim \ga_1 \ga_2 ... \ga_n$, has been proven to be 
incompatible with the Ehrenfest principle, see Eq. (10) below. }     
\beq 
\what{p}{}_a^\mu= -i\hbar\kappa\gamma^\mu\der_a, 
\quad \what{\om}{}_\mu = - 1/\kappa \gamma_\mu  
\eeq 
if the law of composition of operators 
when calculating the  commutator 
corresponding to (2c) is assumed to be the 
symmetrized Clifford (=matrix) product. 
Note that the coefficient $\kappa$ in (3) has the dimension 
of ${\em length}^{-(n-1)}$ and its value is expected to be 
``very large'' in order 
to 
agree with the infinitesimal 
character 
of the $(n-1)$-volume element $\om_\mu$. A possible relation of 
$\kappa$ to the ultra-violet cutoff scale is discussed in 
\cite{qs96,inprep}. 

The realization of operators in terms of 
Clifford imaginary units  
implies a 
certain 
generalization of the formalism of quantum mechanics. 
Namely, whereas  the conventional quantum mechanics 
is built up on complex numbers    
which are  
essentially the Clifford numbers corresponding to the 
{\em one}-dimensional space-time (= the time dimension in mechanics),  
 the present  approach to  quantization of fields 
viewed as 
multi-parameter  Hamiltonian systems 
(of the De Donder-Weyl type) 
makes use of  the hypercomplex (Clifford) algebra of 
the corresponding space-time manifold\cite{bt,hestenes}. 

This philosophy and other considerations based essentially on 
the correspondence principle  
enable  us to  
conjecture the following generalization of the Schr\"odinger equation 
to 
the present framework 
\beq 
i\hbar\kappa \ga^\nu \der_\nu \Psi = \what{H}  \Psi , 
\eeq 
in which  the components of the wave function $\Psi$ are  functions 
of the space-time and field variables: $\Psi=\Psi(x^\mu,y^a)$. 
If $\Psi$ is  a Dirac spinor 
and $\what{H} $ is Hermitian with respect to the  $L^2$ scalar 
product in  $y$-space,   
Eq.~(4)   possesses a positive definite 
conserved quantity as a consequence of the conservation law 
\beq
\frac{\delta}{\delta \sigma} \int_\sigma \omega_\mu \int dy {\Psib}   
\gamma^\mu \Psi =0, 
\eeq
where ${\Psib}$ denotes the Dirac conjugate of $\Psi$, 
and $\sigma$ is any space-like hypersurface.  This 
makes possible a probabilistic interpretation of $\Psi$.  

In the case of a system of scalar fields $\{ y^a \}$ 
given by the Lagrangian 
\beq
 L= \half \der_\mu y^a \der^\mu y_a - V(y)
\eeq
the polymomenta and 
the classical DW Hamiltonian read 
\beq
p^a_\mu= \der_\mu y^a,
 \quad 
H= \half p^a_\mu p_a^\mu + V(y). 
\eeq 
The corresponding quantum counterpart of $H$ 
is shown in 
\cite{qs96,bial97}  
to be 
\beq
 \what{H} = -\half \hbar^2 \kappa^2 \der_a \der^a + V(y) .
\eeq 
Note that for  
a single free scalar field $V(y)=(1/2\hbar^2) m^2 y^2$, 
so that $\what{H}$ 
is similar to the Hamiltonian operator of the harmonic oscillator 
in the space of field variables.  

One of the reasons to believe that (4) is 
a reasonable candidate to a proper 
wave equation is that 
it is consistent with the Ehrenfest 
principle. By assuming\footnote{It should be noticed that the scalar 
product $\int dy \Psib\Psi$  is not positive definite if $\Psi$ is a 
Dirac spinor. However, another version\cite{inprep} of 
the Ehrenfest theorem can also be proven 
using the positive definite scalar product implied by (5). }  
\beq
\left < \what{O}\right > :=\int dy \Psib\what{O}\Psi
\eeq 
{}from equations (3), (4) and (8) we obtain\cite{bial97}   
\beqa 
\der_\mu \left < \what{p}{}_a^\mu\right >  
&=& - \der_\mu \int dy \overline{\Psi}
 i  \hbar \kappa \ga^\mu \der_a{\Psi} 
=...= - \left < \der_a \what{H}\right >    
\nn \\ 
\der_\mu\left < \widehat{y_a \omega^\mu}\right >  
&=& - \kappa^{-1} \der_\mu \int dy \overline{\Psi}\ga^\mu y_a  \Psi 
=...= \left < \what{p^\mu_a \omega_\mu}\right >  . 
\eeqa 
By comparing the result with (1) we conclude that 
the 
classical field 
equations are fulfilled ``in average'' as a consequence of the 
representation of operators (3), 
the generalized 
Schr\"odinger equation (4),  
and  formula (9) for the expectation values of operators. 
Note also that using an appropriate (hypercomplex) analogue of the 
quasiclassical ansatz for the wave function $\Psi$ 
it is possible to derive\cite{qs96,bial97} 
{}from (4) and (8) the Hamilton-Jacobi 
equation corresponding to the DW canonical theory\cite{rund}. 
%
%

\section{Generalization to curved space-time} 

Our next  step is to extend the generalized 
Schr\"odinger equation (4) to  curved space-time given 
by a metric tensor $g_{\mu\nu}(x)$.  
To do this we have 
(i) to introduce 
$x$-dependent $\gamma$-matrices $\gamma^\mu (x)$ such that 
$$
\gamma_\mu(x) \gamma_\nu(x) + \gamma_\nu(x) \gamma_\mu(x)
=2 g_{\mu\nu}(x), 
$$
(ii) to replace the partial 
space-time derivative in the left hand side of (4) with 
the covariant derivative 
$\nabla_\mu$, and 
(iii) to generalize the operator $\what{H}$ to 
a curved space-time background.  
Denoting $g:={\rm det}||g_{\mu\nu}||$ 
the resulting generalized Schr\"odinger equation in 
curved space-time  assumes  the form  
\beq
i\hbar\kappa \sqrt{|g|} \gamma^\mu (x) \nabla_\mu \Psi 
= \what{H} \Psi .  
\eeq 

For the system of scalar fields $\{ y^a \}$ 
on 
a  
curved background we obtain\cite{inprep} 
\beq 
\what{p}{}^\mu_a = - i\hbar\kappa \sqrt{|g|} \gamma^{\mu}\der_a, 
\quad 
\what{H} = -\frac{\hbar^2\kappa^2}{2}
\sqrt{|g|}  \frac{\der^2}{\der y^a \der y_a} 
+ \sqrt{|g|} V(y) . 
\eeq 

If the  wave function $\Psi$ in (11) 
is a Dirac spinor 
the covariant derivative 
$\nabla_\mu:= \der_\mu + 
\theta_\mu$ includes 
the spinor connection   
\beq 
\theta_\mu = \half \theta_{AB}{}_\mu \gamma^{AB}, 
\eeq 
where  
$\gamma^{AB}:=\half( \gamma^A \gamma^B - \gamma^B\gamma^A), 
\quad 
\gamma^A \gamma^B + \gamma^B\gamma^A = 2 \eta^{AB}, 
$ 
 and $\eta^{AB}$ is the  
Minkowski metric.  
The coefficients of the spinor connection are 
given by 
\beq
\theta^A{}_{B\mu} = -e^\nu_B \der_\mu e^A_\nu 
+ e^A_\al \Ga^\al{}_{\mu\nu} e^\nu_B , 
\eeq 
where the tetrad coefficients $e_A^\mu$ 
satisfy 
\beq
g^{\mu\nu}(x)= e_A^\mu (x) e_B^\nu (x) \eta^{AB}  ,  
\eeq
as usual. 
Note that 
   an  analogue of the Ehrenfest theorem 
also  holds here, at least for scalar fields,  
due to the 
well-known relation  for the spinor connection: 
$\der_\mu(\sqrt{|g|} e^\mu_A)
= \sqrt{|g|} e^\mu_B \theta^B{}_{A\mu}$  
(the proof is to be presented elsewhere\cite{inprep}).

\section{Application to General Relativity}

Now we can sketch how the present framework can be applied 
to General Relativity. The wave function will depend 
on the metric (or tetrad) and the space-time variables, 
i.e. $\Psi= \Psi (x^\mu, g^{\al\be}) $ 
(or 
$\Psi=\Psi(x^\mu, e^\al_A)$).  
Here  the metric components $g^{\al\be}$ are not functions 
of space-time variables as in Sec. 3. They enter 
rather as  fibre coordinates of the bundle of symmetric 
rank two tensors over the space-time,    
with classical metrics $g^{\al\be}(x)$ being  sections in this bundle.  
With $g^{\mu\nu}=: e_A^\mu e_B^\nu\eta^{AB}$ 
we introduce 
 $\gamma$-matrices  
$\gamma^\mu:= e_A^\mu\gamma^A $
which fulfill 
$$
\gamma_\mu \gamma_\nu + \gamma_\nu \gamma_\mu
= 2 g_{\mu\nu} .  
$$ 

Now, equations (4) and (11) in the case of quantum  
General Relativity generalize to 
\beq
i\hbar\kappa 
\sqrt{|g|} 
\gamma^\mu \what{\nabla}{}_\mu \Psi = \what{H} \Psi  ,   
\eeq 
where   
$\what{H}$ is the operator of the DW Hamiltonian density  
 and $\what{\nabla}_\mu$ denotes the operator corresponding to the 
covariant derivative in which the connection coefficients are 
replaced by appropriate differential operators (cf. Eq. (20)).  
Note 
that in gravitation theory  it 
seems very natural to 
relate the 
constant $\kappa$ with a Planck scale quantity, so that 
we can expect $\kappa\sim ${\em l}${}_{Pl}^{\,-(n-1)}$.

In order to construct  operators 
entering  Eq.~(16) we need to 
represent 
General Relativity in 
a DW-like Hamiltonian form. 
This issue has already been considered by several 
authors\cite{gimm,esposito}.   
However,  the formulation given by 
Ho\v rava\cite{horava} seems to be  
the  most 
suitable 
for our purposes.   
 Using the metric density 
$ h^{\alpha\beta}:= \sqrt{|g|}g^{\al\be}$ 
as a field variable and 
by defining the quantity
\beq 
 8\pi G\,   
Q^\al_{\beta\gamma} := 
\half[\delta^\alpha_\beta\Gamma^\delta_{\gamma\delta} 
+ \delta^\alpha_\gamma\Gamma^\delta_{\beta\delta} ]
  - \Gamma^\al_{\beta\gamma} ,  
\eeq 
and the DW Hamiltonian density 
\beq
H(h^{\alpha\be}, Q^\al_{\beta\gamma}) :=  
 8\pi G\, 
h^{\alpha\ga} [ Q^\de_{\al\be}Q^\be_{\ga\de}+ 
\frac{1}{1-n}\, Q^\be_{\al\be}Q^\de_{\ga\de} ] , 
\eeq  
the Einstein field equations can be 
written 
in 
the 
DW Hamiltonian form  (cf. Eq.~(1)) 
\beq    
\der_\al h^{\be\ga}
= 
\der H / \der Q^\al_{\be\ga} , 
\quad 
\der_\al Q^\al_{\be\ga}
= 
- \der H / \der h^{\be\ga}  . 
\eeq  
The first of Eqs. (19) is equivalent  to the well-known expression 
of the Christoffel symbols in terms of the metric. 
The second one yields the 
   	vacuum 
Einstein equations 
   	$R_{\al\beta}(\Gamma)=0$  
for the 
Christoffel symbols. 
Note that 
variables $Q^\al_{\beta\gamma}$ 
play the role of polymomenta associated with the 
space-time derivatives of 
the 
metric density  $h^{\alpha\be}$.

Now, the already familiar canonical brackets 
for forms constructed {}from  
polymomenta $Q^\alpha_{\beta\gamma}$ 
and field variables $h^{\mu\nu}$ can be written down 
(cf. Eqs. (2)) and quantized according to Eq.~(3). 
This  
results in 
the following  
expression for the operator 
$\what{Q}{}^\alpha_{\beta\gamma}$: 
\beq
\what{Q}{}^\alpha_{\beta\gamma} = 
-i\hbar \kappa 
 \sqrt{|g|} 
\gamma^\alpha  
\frac{\der}{\der h^{\beta\gamma}} . 
\eeq 
Substituting  (20) to the classical expression (18) 
we obtain the following 
operator of the DW Hamiltonian density\footnote{Recall that in 
$n$ dimensions the Newton gravitational constant 
$G\sim l_{Pl}^{n-2}/\hbar$. }  
\beq
\what{H} 
= - 8\pi G\, 
\hbar^2\kappa^2 \frac{n-2}{n-1}  
\left \{  \sqrt{|g|} h^{\al\ga}h^{\be\de}\frac{\der}{\der h^{\al\be}} 
\frac{\der}{\der h^{\ga\de}} \right \}_{ord} ,    
\eeq 
where the notation $\left \{ ... \right \}_{ord}$ 
refers to the ordering ambiguity 
of the expression inside the curly brackets. 
In fact, the form of $\what{H}$ 
as written down in (21) corresponds to the 
arbitrarily chosen ``standard'' ordering 
when all differential operators appear on the right hand side.  
We hope that by  requiring  the Hermicity of $\what{H}$ 
and applying  the correspondence principle 
we can chose a proper  
ordering of operators in (21)  and  obtain 
a 
more precise expression for 
        the DW Hamiltonian operator 
of gravity (work in progress). 

Further, if the wave function in (16) is assumed  to be 
a spinor,   the operator of the covariant derivative 
$\what{\nabla}_\mu$ 
is 
$\what{\nabla}_\mu:= 
\der_\mu + \what{\theta}_\mu$,  
where 
$\what{\theta}_\mu$ 
denotes the operator corresponding to  the spinor connection.  
Classically the 
spinor connection is given by Eqs. (13-14).   
{}{}From (14) it is clear   that the metric formulation 
does not allow us to construct the operator 
$\what{\theta}_\mu$,  because  
it  includes    the operator corresponding to the 
first jet components (=derivatives) of tetrads 
$\der_\mu e^A_\nu$.  
This operator can be derived only {}from 
 the 
quantization of 
polymomentum variables  
associated with the derivatives of tetrads. 
We thus arrive at the problem of a 
DW-like formulation of General Relativity 
in tetrad variables (work  in progress).   
This formulation  will, of course, lead to another expression for the 
DW Hamiltonian density in terms of tetrads and corresponding 
polymomenta and, consequently, to a different {}from (21) 
expression for the  DW Hamiltonian operator.  
The consistency with the Ehrenfest principle can be checked 
only 
if the operator corresponding to the spinor connection is found.


\section{Discussion} 

In this communication we have pointed out the approach 
to the quantization of fields and general relativity which 
emanates {}from the 
manifestly covariant generalization of the Hamiltonian formalism 
to field theory known as the De Donder-Weyl theory. 
Within the latter fields are treated as 
multi-parameter  De Donder-Weyl Hamiltonian systems  
described by Eqs. (1),   
with  space-time variables entering as parameters 
similar  to the single time parameter in the usual Hamiltonian 
formalism.   The corresponding quantization  
        involves the space-time Clifford algebra 
as a higher-dimensional 
generalization of the algebra of complex numbers 
used in quantum mechanics; 
in doing so the standard quantum   mechanics 
is derived 
as a special case 
corresponding to the one-dimensional ``field theory''.   
Notice that some of our results 
do not necessarily require $\gamma^\mu$-s 
in the above formulas 
to be the Dirac matrices. 
However, the 
question whether or not other hypercomplex systems 
appearing 
in various first order relativistic wave equations\cite{corson} 
could be suitable for the present framework has not been studied 
by us. 

The description of quantized fields is achieved here in terms of 
partial differential equations and functions over a finite 
dimensional covariant analogue of the configuration space --  
the space of field and space-time variables. This description 
appears to be very different as compared with, say, the Schr\"odinger 
functional representation in quantum field theory which employs 
an infinite dimensional configuration space, functionals, and functional 
derivative equations. Still, a qualitative argument based on the 
physical meaning of our wave function 
and 
of 
the Schr\"odinger 
wave functional suggests that a connection between both can exist 
(see\cite{qs96} for  preliminary discussion): 
while our wave function $\Psi(y,\bx,t)$ is 
the probability amplitude that a field  has the value $y$ 
in the space point $\bx$ in the moment of time $t$,  
the Schr\"odinger functional $\Psi([y(\bx)],t)$ is 
the probability amplitude that a field has  the configuration 
$y(\bx)$ in the moment of time $t$. Hence, the Schr\"odinger functional 
could in principle be related to a certain composition of single 
amplitudes given by our wave function. 

There are several 
open questions regarding the generalized 
Schr\"odinger equations ~(4),~(11),~(16).  
Particularly, it is not quite clear yet whether or not the 
wave function $\Psi$ should be a (universal) Dirac spinor for 
any field (of any spin) under quantization. 
If we accept the idea of a universal spinor wave function in (4) 
the question naturally 
arises as to how this can 
be reconciled 
with the spin content of 
different fields we are to quantize. 
This problem is similar to 
the early attempts to construct  matter {}from 
a universal spinor field, 
such as the Heisenberg nonlinear spinor theory and De Broglie's 
``neutrino theory of light'', which are nowdays abandoned.  

On the one hand, 
a possible way out 
may be 
that equation (4) (and  (11),~(16))  should be understood 
rather 
as one of the system of $2s$ equations of 
the Bargmann-Wigner type\cite{corson}   
for 
a {\em multi\,}spinor wave function 
with $2s$ (antisymmetric) indices,  
which describes particles of spin $s\geq 1/2$   
(the case of 
a 
scalar field would have to be given 
a separate treatment then). However,  it is not clear to us 
how this point of view  can be extended to systems like QED 
which involve interacting fields of different spins.   

On the other hand, 
if 
$\Psi$ 
is taken to be  
a general element of the Clifford algebra 
its  component content  may probably  
suffice to incorporate  fields of different 
spins (including spinors if  
viewed as minimal left ideals\cite{bt}). 
In this light,  
our earlier assumption\cite{qs96} 
that $\Psi$ is a hypercomplex valued function  
(or a non-homogeneous differential form\cite{bial94}), 
which we have abandoned in\cite{bial97} 
because of 
difficulties (which probably  can be surmounted)  
with the probabilistic interpretation 
and the Ehrenfest principle, 
could still be worthy of further analysis.  
Moreover, when passing on to  curved space-time 
the use of (multi)spinor wave functions restricts the 
applicability of the present framework to 
space-times 
which admit  spinor structure. 
Such restriction seems to be rather artificial,     
especially 
if we think 
about a subsequent application 
to quantization of General Relativity. By  requiring 
that no restriction to spin space-time manifolds is allowed 
we arrive at yet another alternative 
by using  the Dirac-K\"ahler 
equation\cite{bt,dk} rather than the Dirac spinor equation.   

Let us recall that in Minkowski space-time the Dirac-K\"ahler equation 
leads to 
the same predictions as the Dirac equation. 
However, in curved space-time it exhibits different behavior 
which is incompatible with the Pauli exclusion principle\cite{bt}.  
This feature, however, cannot be treated as a drawback if the 
Dirac-K\"ahler equation is used 
not to describe the electron with spin but as 
as a more fundamental equation.   

The deep geometric content of the Dirac-K\"ahler equation reconciles 
remarkably  with the geometric spirit of General Relativity.  
On this ground the Dirac-K\"ahler version of (16) 
can be put forward as a 
better (although still hypothetic) 
candidate for description of  quantum General Relativity
\beq 
i\hbar\kappa \sqrt{|g|}\, 
\what{\mbox{$\hspace*{0.3em}\not\mbox{\hspace*{-0.3em}D}$}_\Gamma} 
\Psi = \what{H} \Psi . 
\eeq
Here 
${\mbox{$\hspace*{0.3em}\not\mbox{\hspace*{-0.3em}D}$}_\Gamma} := 
d_\Gamma -\delta_\Gamma $ 
is the Dirac-K\"ahler (Hodge-de Rham) 
covariant differential operator 
related to  the metric connection $\Gamma$, with 
the hat over it in (22) denoting  that the connection coefficients 
$\Gamma^\al_{\be\ga}$ are 
differential operators  themselves (cf. Eq.~(20)). 
The wave function in (22) is a non-homogeneous form: 
$\Psi=\sum_{p=0}^{n}\frac{1}{p!}
\psi\lind{\mu}{p}dx^{\mu_1}\we ...\we dx^{\mu_p}$,    
and the operators have to be represented using the elements of 
the Atiyah-K\"ahler algebra (e.g. $\der_\mu \inn$ and $dx^\mu \we$) 
rather than Dirac matrices. 
In this formulation we already do not need to 
employ the tetrad formulation of gravity. 
All the quantities can in principle be determined {}from 
quantization of 
metric gravity in the DW formulation. 
Potential difficulties 
which may emerge again   
{}from the operator ordering ambiguity 
probably can be 
overcome with the help of the correspondence  principle. 

Further work will hopefully tell us which of the  
alternatives outlined above is more consistent;   
what is the physical content of the presented formalism, 
and how it can be linked to the contemporary 
quantum field theory. 


When the paper was being written 
the preprint by 
M. Navarro\cite{nav} has appeared 
in which he discusses an approach to a ``finite dimensional''  
quantization of fields  similar to ours in Sect. 2. 


\end{document}